\documentclass[prl,aps,amsfonts,amssymb,amsmath,floats,floatfix,twocolumn,superscriptaddress,unsortedaddress]{revtex4}
\usepackage{graphics}
\usepackage{epsfig,graphicx}

\begin{document}

\title{Loss of nodal quasiparticle integrity in underdoped YBa$_2$Cu$_3$O$_{6+x}$}


\author{D. Fournier}
\affiliation{Department of Physics {\rm {\&}} Astronomy, University of British Columbia, Vancouver, British
Columbia V6T\,1Z1, Canada}
\author{G. Levy}
\affiliation{Department of Physics {\rm {\&}} Astronomy, University of British Columbia, Vancouver, British
Columbia V6T\,1Z1, Canada}
\author{Y. Pennec}
\affiliation{Department of Physics {\rm {\&}} Astronomy, University of British Columbia, Vancouver, British
Columbia V6T\,1Z1, Canada}
\author{J.L. McChesney}
\affiliation{Advanced Light Source, Lawrence Berkeley National Laboratory, Berkeley, California 94720, USA}
\author{A. Bostwick}
\affiliation{Advanced Light Source, Lawrence Berkeley National Laboratory, Berkeley, California 94720, USA}
\author{E. Rotenberg}
\affiliation{Advanced Light Source, Lawrence Berkeley National Laboratory, Berkeley, California 94720, USA}
\author{R. Liang}
\affiliation{AMPEL, University of British Columbia, Vancouver, British Columbia V6T\,1Z4, Canada}
\author{W.N. Hardy}
\affiliation{Department of Physics {\rm {\&}} Astronomy, University of British Columbia, Vancouver, British
Columbia V6T\,1Z1, Canada} \affiliation{AMPEL, University of British Columbia, Vancouver, British Columbia
V6T\,1Z4, Canada}
\author{D.A. Bonn}
\affiliation{Department of Physics {\rm {\&}} Astronomy, University of British Columbia, Vancouver, British
Columbia V6T\,1Z1, Canada} \affiliation{AMPEL, University of British Columbia, Vancouver, British Columbia
V6T\,1Z4, Canada}
\author{I.S. Elfimov}
\affiliation{AMPEL, University of British Columbia, Vancouver, British Columbia V6T\,1Z4, Canada}
\author{A. Damascelli}
\affiliation{Department of Physics {\rm {\&}} Astronomy, University of British Columbia, Vancouver, British
Columbia V6T\,1Z1, Canada} \affiliation{AMPEL, University of British Columbia, Vancouver, British Columbia
V6T\,1Z4, Canada}

\maketitle

{\bf Arguably the most intriguing aspect of the physics of cuprates is the close proximity between the record high-$T_c$ superconductivity
(HTSC) and the antiferromagnetic charge-transfer insulating state driven by Mott-like electron correlations. These are responsible for the
intimate connection between high and low-energy scale physics \cite{meinders,kohsaka,darren}, and their key role in the mechanism of HTSC was
conjectured very early on \cite{anderson}. More recently, the detection of quantum oscillations in high-magnetic field experiments on
YBa$_2$Cu$_3$O$_{6+x}$ (YBCO) has suggested the existence of a Fermi surface of well-defined quasiparticles in underdoped cuprates
\cite{Doiron2007,sebastianCM}, lending support to the alternative proposal that HTSC might emerge from a Fermi liquid across the whole cuprate
phase diagram \cite{comanac,sachdev}. Discriminating between these orthogonal scenarios hinges on the quantitative determination of the elusive
quasiparticle weight $Z$, over a wide range of hole-doping $p$. By means of angle-resolved photoemission spectroscopy (ARPES) on in situ doped
YBCO \cite{hossain}, and following the evolution of bilayer band-splitting, we show that the overdoped metal electronic structure
($0.25\!\lesssim\!p\!\lesssim0.37$) is in remarkable agreement with density functional theory \cite{andersen,elfimov,pasani} and the
$Z\!=\!2p/(p\!+\!1)$ mean-field prediction \cite{andersonvanilla,yangrice}. Below $p\!\backsimeq\!0.10$-$0.15$, we observe the vanishing of the
nodal quasiparticle weight $Z_N$; this marks a clear departure from Fermi liquid behaviour and -- consistent with dynamical mean-field theory
\cite{gabriel} -- is even a more rapid crossover to the Mott physics than expected for the doped resonating valence bond (RVB) spin liquid
\cite{andersonvanilla,yangrice}.}

Formally, the degree of quasiparticle integrity is revealed by $Z_k\!\equiv\!\int\! A_{coh}(k,\omega)\,d\omega$, i.e. the integrated spectral
weight of the coherent part of the single-particle spectral function $A(k,\omega)\!\equiv\!A_{coh}(k,\omega)\!+\!A_{incoh}(k,\omega)$ probed by
ARPES \cite{Damascelli:RMP,Campuzano}. Experimentally, while in the optimally-to-overdoped regime $Z_k$ is believed to be finite -- yet
quantitatively undetermined -- at all momenta both above and below $T_c$, the situation is much more controversial in the
optimally-to-underdoped regime \cite{Damascelli:RMP,Campuzano}. Although it has been conjectured that the $T\!=\!0$ extrapolation of the
pseudogap state is a nodal (N) liquid \cite{kanigel,chatterjee}, a determination of $Z_N$ has not been possible. As for the antinodal (AN)
quasiparticle spectral weight, the ARPES spectra are characterized by a dramatic temperature dependence, with broad incoherent features in the
pseudogap state and quasiparticle-like excitations emerging below $T_c$ \cite{Damascelli:RMP,Campuzano}; this observation led to the early
proposal that the onset of HTSC might be thought of as a `coherence transition' \cite{zxgeorge}. Detailed doping-dependent studies reported a
decreasing $Z_{AN}$ upon reducing $p$ \cite{donglai,ding}; however, whether $Z_{AN}\!\varpropto\!p$ \cite{ding} or vanishes at finite $p$
\cite{donglai} remains unresolved, because of the experimental difficulty in the quantitative discrimination between $A_{coh}$ and $A_{incoh}$.
As discussed in greater detail in the Supplementary Information, where we present an analysis of the quasiparticle spectral weight across the
whole YBCO phase diagram, this is particularly challenging on the underdoped side where $Z_k$ vanishes; and even in the overdoped regime, this
allows at\,best a relative -- rather than absolute -- determination of $Z_k$.

As an alternative, potentially more quantitative approach, in materials with CuO$_2$ bilayers within the unit cell such as
Bi$_2$Sr$_2$CaCu$_2$O$_{8+\delta}$ (Bi2212) and YBCO, the quasiparticle strength $Z_k$ might be estimated from the bonding (B) and antibonding
(AB) band splitting $\epsilon^{B,AB}(k)\!=\!\epsilon(k)\mp t_{\perp}(k)$, where $\epsilon(k)$ is the quasiparticle dispersion with respect to
the in-plane momentum $k$, and $t_{\perp}(k)$ accounts for the interplane coupling.
\begin{figure}[t!]
\includegraphics[width=1\linewidth]{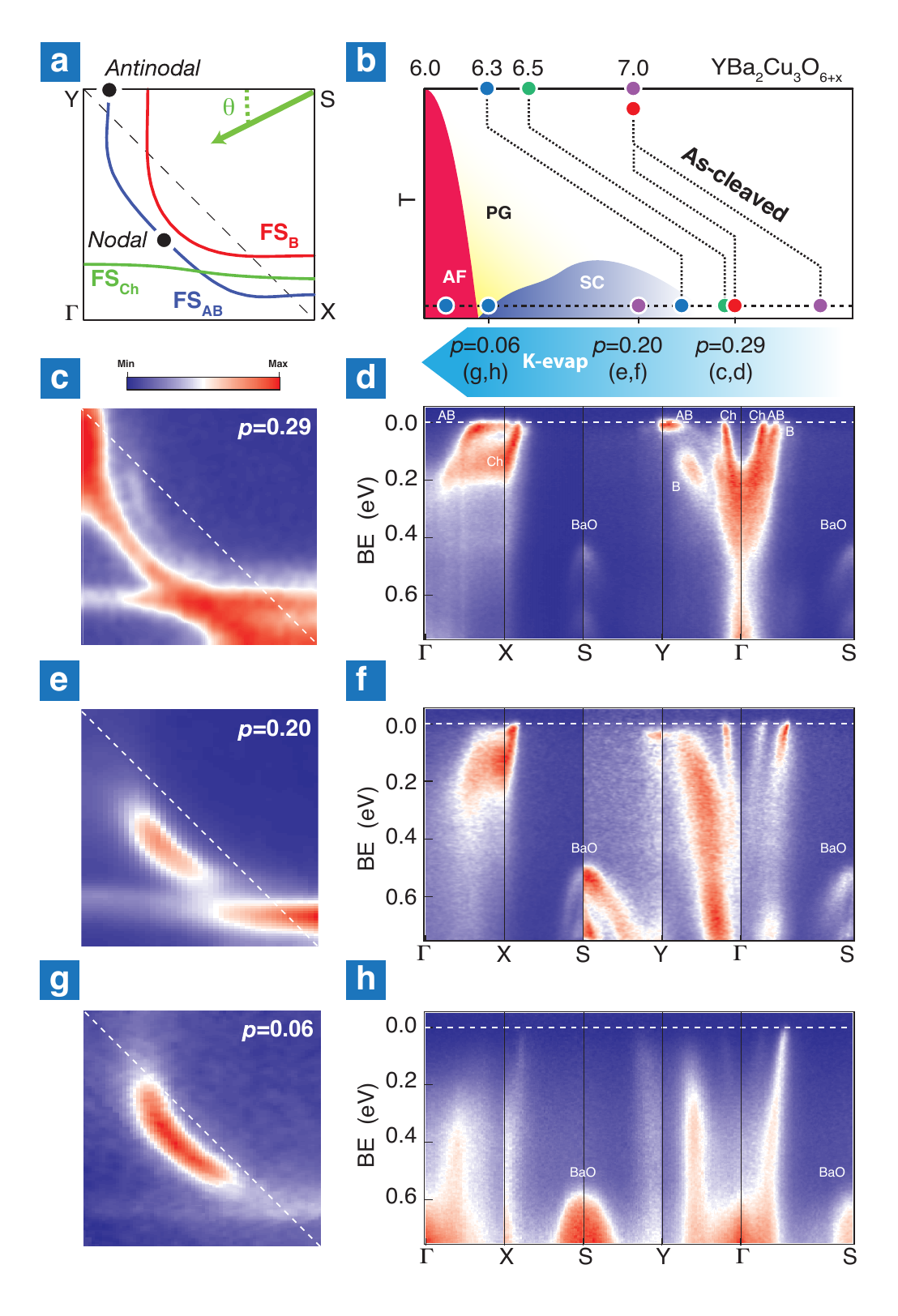}
\caption{{\bf Fermi surface and band dispersion across the YBCO phase diagram}. (a) Schematics of the Fermi
surface of optimally doped YBCO from band-structure calculations: three sheets of Fermi surface ($FS_{Ch}$,
$FS_{B}$, $FS_{AB}$) are derived from the one-dimensional CuO-chain (Ch) band, and the bonding (B) and
antibonding (AB) CuO$_2$-plane bands. (b) YBCO doping-range for bulk (oxygen content, top axis) and surface
(hole-doping $p$, bottom axis) investigated in this work. Color dots refer to different samples of various
oxygen content; from the combination of as-cleaved and K-deposited surfaces, a doping $p$ ranging from 0.37
to 0.02 could be accessed. (c-h) Fermi surface and band dispersion, as determined by ARPES at $T\!=\!20$\,K,
for three representative hole-doping levels obtained on: self-doped as-cleaved YBCO7 [(c,d), $p\!=\!0.29$,
red dot in b], a second K-evaporated YBCO7 sample [(e,f), $p\!=\!0.20$, purple dot in b], and heavily
K-evaporated YBCO6.3 [(g,h), $p\!=\!0.06$, blue dot in b]. At $T\!=\!20$\,K the data are representative of
the overdoped normal metal (c, large $FS_{B}$ $FS_{AB}$ barrels), the nearly optimally doped superconductor
(e, gapped Fermi surface with nodal Dirac points), and the underdoped pseudogap state (g, Fermi
arcs).}\label{NSevolution}
\end{figure}
Access to $Z_k$ is provided by the fact that both $\epsilon(k)$ and $t_{\perp}(k)$, as measured by ARPES,
correspond to {\it effective} quantities; under the assumption that correlation effects renormalize the
bandwidth but {\it not} the functional form of the quasiparticle dispersion, these effective quantities are
related to the {\it density functional theory} values through the same renormalized quasiparticle strength
$Z_k$. This assumption is most appropriate along the nodal direction of cuprates
\cite{Damascelli:RMP,Campuzano}, which is the focus of our study since {\it nodal quasiparticles} -- being
unaffected by the opening of pseudo and superconducting gaps -- provide access to the normal state underlying
the emergence of HTSC. This approach for the extraction of $Z_k$ will be experimentally validated based on a
direct comparison with results for the quasiparticle spectral weight.

The renormalization of the in-plane hopping probability due to correlations is given by the ratio between the
number of ways an electron can be added to the Mott-like system as compared to the independent particle
system. While in the latter this is proportional to the total number $1+p$ of available band states [1 at
half-filling ($p\!=\!0$), and 2 for a completely empty band ($p\!=\!1$)], in the Mott-correlated case it is
proportional to the doping $p$ away from half-filling, times a factor of 2 because for each hole induced by
doping an electron can be added with spin up or spin down \cite{meinders}. This leads to the mean-field
result $Z\!=\!2p/(p\!+\!1)$ \cite{andersonvanilla,yangrice}, and is the same for in-plane and interplane
hopping since both processes are governed by the same correlation effects. The effective bilayer splitting
can be written as $\Delta\epsilon^{B,AB}_k \!=\! 2t_{\perp}(k)\!=\! 2 Z_k t_{\perp}^{LDA}(k)$; its detection
by ARPES and the a-priori knowledge of $t_{\perp}^{LDA}(k)$, as derived from local-density approximation
(LDA) band structure calculations, allows a quantitative determination of $Z_k$.

A study of this kind has been performed on Bi2212 at the antinodes in the optimal-to-overdoped regime
\cite{adam}; however, a systematic investigation in the underdoped regime is lacking. A potentially more
promising candidate for such a study of $Z_k$ is YBCO, which is characterized by a larger bilayer splitting
especially along the nodal direction \cite{andersen,elfimov,pasani}; furthermore, although YBCO has long been
considered unsuitable for ARPES because of the polar-catastrophe-driven overdoping of the cleaved surface
\cite{hossain,pasani}, the in situ doping method based on the deposition of potassium allows exploration of a
doping range much wider than for any other cuprate family \cite{hossain}. Here we apply this approach to
detwinned YBa$_2$Cu$_3$O$_{6+x}$ single crystals for several values of the bulk oxygen content:
$(6\!+\!x)\!=\!6.34$, 6.35, 6.51, and 6.99 (hereafter referred to as YBCO6.3, YBCO6.5, and YBCO7). As
summarized in the YBCO phase diagram of Fig.\,\ref{NSevolution}b, by performing ARPES on different as-cleaved
and K-deposited samples we can follow the evolution of the electronic structure from the heavily overdoped
($p\!\simeq\!0.37$, by far the highest among all overdoped cuprates including Tl$_2$Ba$_2$CuO$_{6+\delta}$
\cite{plate}), to the deeply underdoped regime ($p\!\simeq\!0.02$).

In the overdoped regime ($p\!=\!0.29$, Fig.\,\ref{NSevolution}c), we observe three Fermi surfaces ($FS_{Ch}$,
$FS_{B}$, $FS_{AB}$), in agreement with LDA band-structure calculations (Fig.\,\ref{NSevolution}a). These
correspond to the one-dimensional CuO-chain (Ch) band, and the B and AB CuO$_2$-plane bands. The dispersions
of the bands seen by ARPES (Fig.\,\ref{NSevolution}d) are also consistent with the LDA results, with the
exception of the BaO band that in the calculations is located at the Fermi energy \cite{andersen,elfimov},
but in the experiment is found at $\sim\!450$\,meV binding energy (similar disagreement is encountered for
TlO \cite{plate} and BiO bands \cite{linBiO} in Tl- and Bi-cuprates). Upon reaching the underdoped regime by
potassium deposition on the as-cleaved surfaces, behaviour consistent with the established phenomenology of
underdoped cuprates is seen \cite{kanigel,kyle,lee,jeff,hufner}; the CuO$_2$-plane Fermi surface barrels
become partially gapped and reduce to four nodal `Dirac points' (somewhat broadened by resolution) in the
$p\!=\!0.20$ superconductor (Fig.\,\ref{NSevolution}e), and to four extended Fermi arcs in the $p\!=\!0.06$
pseudogap phase (Fig.\,\ref{NSevolution}g). The spectral function becomes progressively more incoherent, with
reduced low-energy quasiparticle intensity and a continuum of spectral weight extending to increasingly
higher binding energy (Fig.\,\ref{NSevolution}f,h). We also observe a progressive change in $FS_{Ch}$ volume
and BaO binding energy (Fig.\,\ref{NSevolution}c-h); as discussed in the Supplementary Information,
monitoring the evolution of these features allows the precise determination of the effective surface
hole-doping $p$ induced by potassium deposition.

\begin{figure}[t!]
\includegraphics[width=1\linewidth]{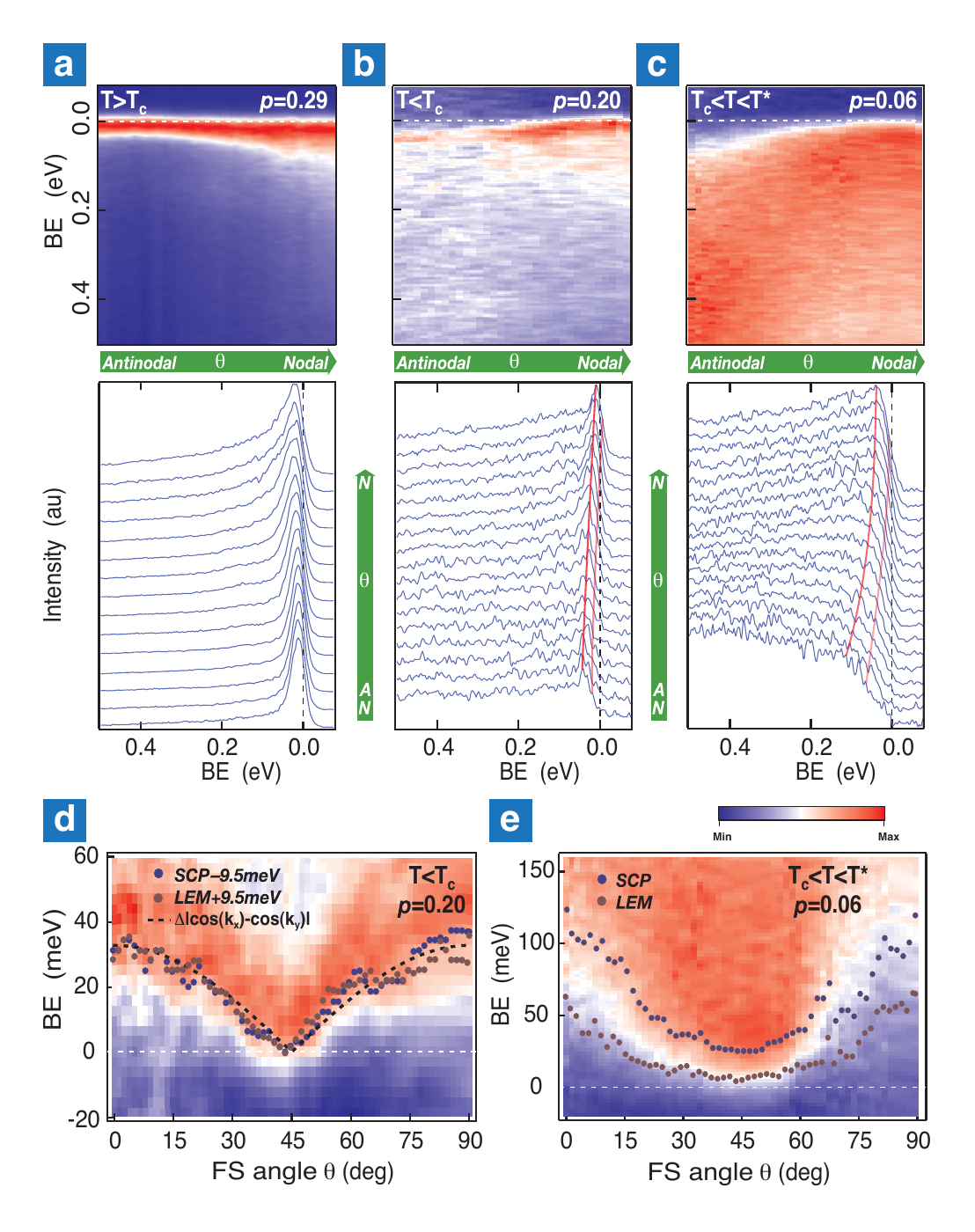}
\caption{{\bf Doping evolution of the gap in YBCO}. (a-c) ARPES image plots and selected energy distribution curves (EDCs), from the antinodal
to the nodal region, along the AB-band Fermi surface and minimum gap contour. The data were obtained at $T\!=\!20$\,K on the same three samples
of Fig.\,\ref{NSevolution}: (a) overdoped normal metal ($p\!=\!0.29$, $T\!>\!T_c$); (b) nearly-optimally doped superconductor ($p\!=\!0.20$,
$T\!<\!T_c$); (c) underdoped pseudogap state ($p\!=\!0.06$, $T_c\!<\!T\!<\!T^*$). While gapless metallic behavior is observed for $p\!=\!0.29$
(a), a v-shape gap consistent with the $d$-wave functional form is detected for $p\!=\!0.20$ (d), as evidenced by the evolution of both
superconducting peak (SCP) and leading-edge midpoint (LEM) as a function of the Fermi surface angle $\theta$ defined in
Fig.\,\ref{NSevolution}a. For $p\!=\!0.06$ (e), a much more rounded gap profile is observed, consistent with the established pseudogap and Fermi
arc phenomenology (in this latter case SCP corresponds to the minima in the second energy derivative of the EDCs).}\label{SCevolution}
\end{figure}

As for the superconductivity related features, Fig.\,\ref{SCevolution} shows image plots and selected energy distribution curves (EDCs) from the
antinodal to the nodal region, along the AB-band Fermi surface and minimum gap contour. The $p\!=\!0.29$ EDCs have no marked momentum
dependence, typical of metallic behavior (Fig.\,\ref{SCevolution}a).
At $p\!=\!0.20$ (Fig.\,\ref{SCevolution}b and d), a clear v-shape gap consistent with the $d_{x^2-y^2}$ HTSC gap is evidenced by the momentum
evolution of both leading-edge midpoint (LEM) and superconducting peak (SCP). The $\Delta\!\mid\! \cos(k_x)\!-\! \cos(k_y)\!\mid$ fit of the
average of the two (Fig.\,\ref{SCevolution}d) returns a maximum antinodal gap value $\Delta\!\simeq\!34$\,meV, consistent with the
$2\Delta\!\sim\!70$\,meV reported for HTSCs with $T_c^{max}\sim95$\,K \cite{hufner}; this also gives $v_{F}/v_2\!=\!15\!\pm\!1$, where $v_2$ is
the slope of the gap at the node and $v_{F}\!=\!1.42\!\pm\!0.05$\,eV\AA. At $p\!=\!0.06$, the quasiparticle peaks and $d$-wave-like
superconducting gap are replaced by broad incoherent features with marginally defined LEM and SCP (Fig.\,\ref{SCevolution}c); these describe a
rounded pseudo-gap-like profile \cite{lee}, with a much larger $\Delta$ at the antinodes (Fig.\,\ref{SCevolution}e). This behavior, and the
associated Fermi arc phenomenology (Fig.\,\ref{NSevolution}g), are hallmarks of the pseudogap phase \cite{kanigel,kyle,lee,jeff,hufner}.

The ($T\!=\!20$\,K) results in Fig.\,\ref{NSevolution} and \ref{SCevolution} reproduce the major features of
the cuprate electronic structure in the overdoped normal metal ($p\!=\!0.29$, $T\!>\!T_c$), the
nearly-optimally doped superconductor ($p\!=\!0.20$, $T\!<\!T_c$), and the underdoped pseudogap state
($p\!=\!0.06$, $T_c\!<\!T\!<\!T^*$). Most importantly, they demonstrate that in situ potassium doping of
cleaved YBCO provides a very effective means of exploring the whole cuprate phase diagram via
surface-sensitive single-particle spectroscopies.
\begin{figure*}[t!]
\includegraphics[width=0.8\linewidth]{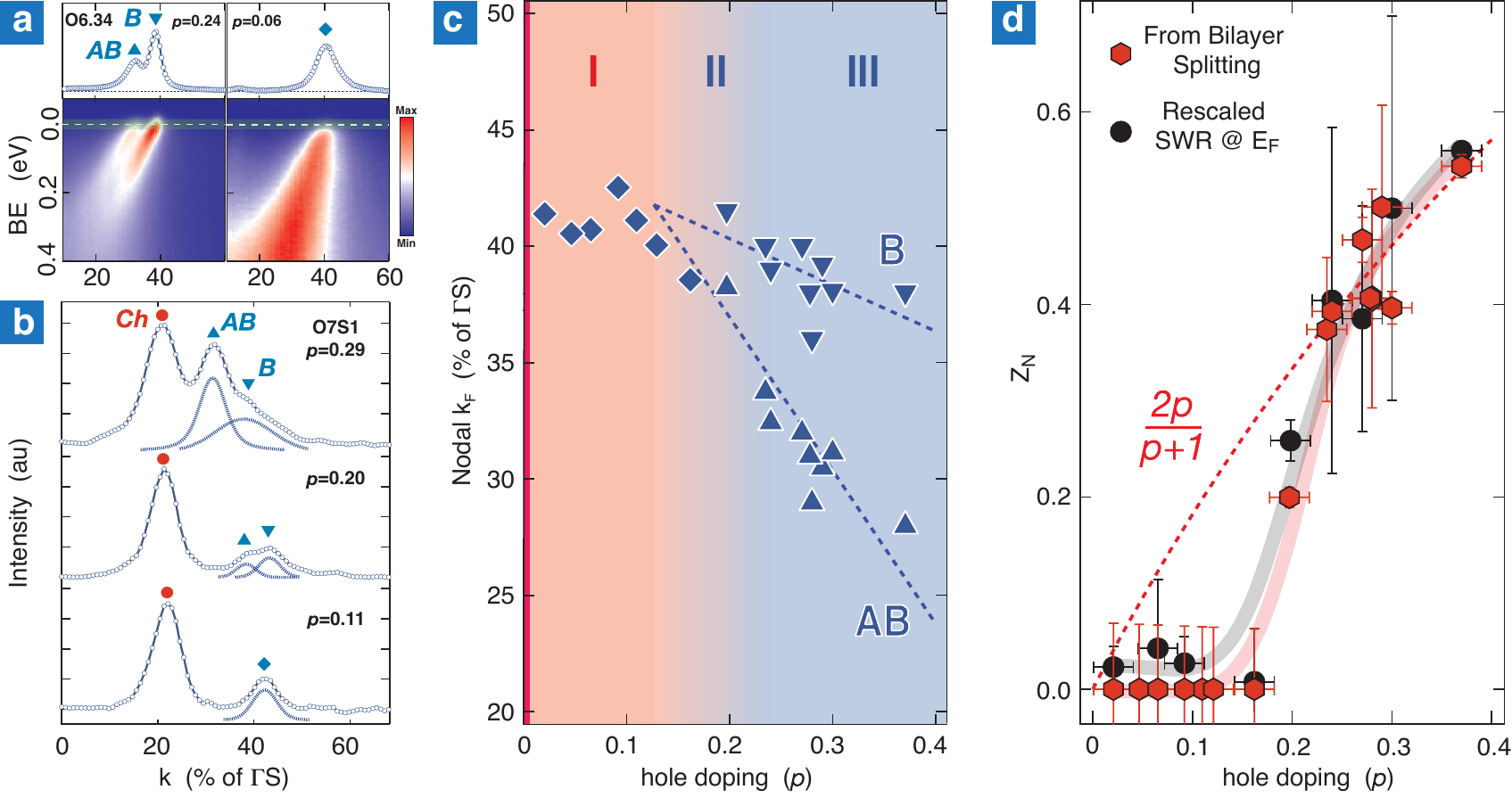}
\caption{{\bf Doping evolution of nodal $k_{F,N}^{B}$, $k_{F,N}^{AB}$, and $Z_N$.} (a) ARPES nodal dispersion
and corresponding $E_F$ momentum distribution curves (MDCs; $\pm15$meV integration, see shaded region), from
as-cleaved and K-deposited YBCO6.34, for light polarization parallel to $\Gamma$-S. (b) Similar MDCs from
YBCO7, for light polarization parallel to $\Gamma$-Y (note the polarization-dependent change of Ch, B, and AB
intensity). (c) Evolution of $k_{F,N}^{B}$ (down triangles) and $k_{F,N}^{AB}$ (up triangles) along the nodal
direction (distance from the $\Gamma$ point in percentage of $\Gamma$-S), as determined by ARPES across the
entire hole-doping phase diagram of YBCO; below $p\!\simeq\!0.15$ the bilayer B-AB splitting vanishes and
only one single $k_{F,N}$ is detected (diamonds). (d) Nodal quasiparticle renormalization $Z_N$ as determined
from the observed bilayer splitting ($Z_N\!=\!\Delta\epsilon^{B,AB}_N/2 t_{\perp}^{LDA}(N)$ with
$t_{\perp}^{LDA}(N)\!\simeq\!120$\,meV) and the low-energy spectral-weight ratio (SWR) defined as
$\int^{-\infty}_{E_F} I(k_{F,N},\omega)\, d \omega / \int^{-\infty}_{0.8\,eV} I(k_{F,N},\omega)\, d \omega$
from the ARPES intensity $I(k,\omega)\!=\!I_0(k)f(\omega)A(k,\omega)$. The SWR was rescaled so that the two
curves match in the $0.23\!-\!0.37$ doping range (see Supplementary Information). Thick gray and orange lines
are spline guides to the eye, while the dashed red line is the mean-field relation $2p/(p\!+\!1)$
\cite{andersonvanilla,yangrice}. For the bilayer splitting data, the error bars are defined from the B-AB MDC
fits when bilayer splitting is detected, and from the experimental resolutions otherwise; for the SWR data,
they are calculated from the spread in the SWR values for integration windows smaller than $[0.1\,{\rm
eV},-\infty]$ (Fig.\,3 in Supplementary Information).}\label{nodalkf}
\end{figure*}
We will now discuss the evolution of $k_{F}$ and B-AB bilayer band splitting along the nodal direction,
aiming at a quantitative determination of $Z_k$. As shown in Fig.\,\ref{nodalkf}a for YBCO6.3, the B and AB
bands are clearly resolved in the overdoped regime ($p\!=\!0.24$), while only one dispersive feature is
detected at low doping values ($p\!=\!0.06$); a similar evolution has been observed on all YBCO samples. By
fitting the momentum distribution curves (MDCs) at $E_F$ with Lorentzian lineshapes convoluted with a
Gaussian to account for the instrumental resolution (Fig.\,\ref{nodalkf}a,b), we extract $k_{F,N}$ from
$p\!=\!0.37$ all the way to $0.02$. Bilayer splitting is progressively reduced upon underdoping and below
$p\!\simeq\!0.15$ only one CuO$_2$ band is detected (Fig.\,\ref{nodalkf}c). From the linear fit of B and AB
nodal $k_{F,N}$ versus doping we identify a vanishing $p_c\!=\!0.12\pm0.02$, consistent with the recent
interpretation of optical -- and thus bulk sensitive -- data from YBCO, which also suggest the suppression of
bilayer splitting between 6.5 and 6.75 oxygen content \cite{dominik}.

As for the evolution of $Z_k$, we start from the heavily overdoped regime ($p\!>\!0.25$, region III in
Fig.\,\ref{nodalkf}c), where ARPES is in overall quantitative agreement with the LDA calculations. At
$p\!=\!0.37$ (i.e., overdoped normal metal), we can accurately determine both the N and AN bilayer
splittings, $\Delta\epsilon^{B,AB}_N\!=\!130\pm10$\,meV and $\Delta\epsilon^{B,AB}_{AN}\!=\!150\pm10$\,meV.
From the LDA values $t_{\perp}^{LDA}(N)\!\simeq\!120$\,meV and $t_{\perp}^{LDA}(AN)\!\simeq\!150$\,meV
(Supplementary Information), and making use of the relation $\Delta\epsilon^{B,AB}_k \!=\! 2 Z_k
t_{\perp}^{LDA}(k)$, we obtain $Z_N\!\simeq\!0.54$ and $Z_{AN}\!\simeq\!0.50$. Upon underdoping, we focus on
the nodal direction where the bilayer splitting can be detected both as $\Delta k_{F,N}^{B,AB}$ and $\Delta
\epsilon^{B,AB}_N$, owing to the presence of the gap nodes. While the results in region II are still at least
qualitatively consistent with LDA ($0.12\!<\!p\!<\!0.25$), the disappearance of bilayer splitting in the
underdoped regime ($p\!<\!0.12$, region I in Fig.\,\ref{nodalkf}c) marks a clear departure from the
independent-particle description. Note that LDA would give a nearly doping-independent $\Delta
k_{F,N}^{B,AB}$ with $(k_{F,N}^{AB}$; $k_{F,N}^{B})\!=\!(33.0;38.7), (35.2;39.5)$, and (38.0;42.2) for O7.0,
O6.5, and O6.0, respectively \cite{elfimov}. Furthermore, the value of bilayer splitting is also rather
insensitive to electrostatic effects associated with the presence of different concentrations of O$^{2-}$
ions in the CuO chain layers or K$^{1+}$ above the BaO terminations, due to the vanishing potential these
charged layers exert on the CuO$_2$ bilayer (Supplementary Information).

The bilayer-splitting-derived results for $Z_N$ across the whole phase diagram are presented in Fig.\,\ref{nodalkf}d, together with the nodal
spectral-weight ratio (SWR) estimated at $E_F$ (see caption for definition). When calculated for integration windows smaller than $[0.1\,{\rm
eV},-\infty]$ in binding energy, the SWR provides a relative measure of the degree of quasiparticle integrity $Z_N$ (Supplementary Information).
Although not a quantitative estimate of $Z_N$, the doping evolution of the low-energy SWR is reminiscent of the quasiparticle renormalization
inferred from the bilayer splitting. When the two results are plotted together (after the necessary rescaling of the SWR, so that the two curves
match in the $0.23\!-\!0.37$ doping range), the similarity is striking (Fig.\,\ref{nodalkf}d), validating the quantitative determination of
$Z_N$ from the evolution of bilayer band splitting. Furthermore, the agreement with the $2p/(p\!+\!1)$ relation above $p\!\simeq\!0.23$ provides
an unprecedented quantitative estimate of quasiparticle integrity in overdoped cuprates, which supports a description of the normal-state
electronic structure based on density functional theory with nearly isotropic quasiparticle renormalization factors, as in the spirit of Fermi
liquid theory. Below $p\!\simeq\!0.23$, however, $Z_N$ deviates from $2p/(p\!+\!1)$, vanishing in the $p\!\backsimeq\!0.10-0.15$ range where the
error bars in Fig.\,\ref{nodalkf}d define an upper limit $Z\!\simeq\!0.065$. From this we can estimate a corresponding upper limit for the nodal
bilayer splitting $\Delta\epsilon^{B,AB}_{N}\!\simeq\!15.6$\,meV, which is consistent with the value of 8\,meV reported based on the analysis of
quantum oscillation results \cite{Dimov,Audouard}. Note also that in region I, not only have the bilayer splitting and $Z_N$ vanished, the
doping evolution of $k_{F,N}$ appears to have saturated, despite the fact that the CuO$_2$ planes are still becoming more underdoped as
evidenced by the further increase of pseudogap magnitude and high-energy incoherent spectral weight. This indicates that, upon approaching the
charge-transfer insulator, the quasiparticle integrity is lost even faster than hinted by renormalized mean-field treatments of the doped spin
liquid \cite{andersonvanilla,yangrice}. It also suggests that the emergence of the Mott physics in cuprates might be captured by cluster
dynamical mean-field theory calculations, which predict a similarly rapid suppression of $Z_N$ in the $p\!\simeq\!0.10\!-\!0.15$ doping range
\cite{gabriel}.

Finally, we note that a vanishing $Z_N$ does not imply vanishing of the total single-particle spectral weight, but rather a redistribution from
coherent to incoherent components of $A(k,\omega)$. While for $p\!=\!0.06$ we find $Z_N\!\simeq\!0$, spectral weight is still present all the
way to $E_F$ (Fig.\,\ref{nodalkf}a and Supplementary Information); however, the distinction between $A_{coh}$ and $A_{incoh}$ is becoming
completely blurred. This suggests something similar to early exact diagonalization cluster calculations for Mott-Hubbard and charge-transfer
models \cite{meinders}, where it is the {\it total} and not just the {\it coherent} spectral weight that vanishes as $2p/(p\!+\!1)$. Indeed,
optical spectroscopy \cite{basov} shows that the total low-frequency ($\omega\!\lesssim\!75$\,meV) integrated spectral weight -- whether
associated with a well defined Drude peak or an incoherent band -- scales to zero with the doping $p$ \cite{basov}. More fundamentally, the loss
of quasiparticle integrity implies the breakdown of concepts such as Fermi surface, band dispersion and Fermi velocity, as well as Luttinger's
sum rule, and suggests that below $p\!\lesssim\!0.10-0.15$ the physical properties of underdoped cuprates (in zero-field) are dominated by
incoherent excitations.

\section{Methods}

\noindent {\bf Sample preparation.} For this study we used detwinned YBa$_2$Cu$_3$O$_{6+x}$ single crystals
with $x\!=\!0.34$, 0.35, 0.51, and 0.99 (see Ref.\,\onlinecite{hossain} for details on growth, annealing, and
detwinning procedures).

\noindent {\bf ARPES experiments.} ARPES measurements were performed at Beamline 7.01 of the Advanced Light Source, with linearly-polarized 100
and 110\,eV photons and a Scienta-R4000 electron analyzer ($\Delta E\!\simeq\!30$\,meV and $\Delta \theta\!\simeq\!0.1^{\circ}$). The light
polarization was parallel to the plane of emission, while the samples were aligned with the in-plane Cu-O bonds either parallel or at
45$^{\circ}$ with respect to the polarization (resulting in more intensity for the antinodal/chain or nodal features, respectively). Fermi
surface maps were obtained by integrating the ARPES intensity over a 30\,meV energy window about $E_F$, and then normalized relative to one
another for display purposes. Mounted on a 5-axis manipulator, the samples were cleaved at 20\,K and 2.5x10$^{-11}$\,torr, and oriented in situ
by taking fast Fermi surface scans.

The self-doping of the YBCO surfaces for various bulk oxygen contents was tuned towards the underdoped regime
by in situ deposition of submonolayers of potassium. We acquired data on the as-cleaved surfaces, as well as
after subsequent potassium evaporations from a commercial SAES alkali-metal dispenser; evaporation current
and time were precisely monitored (in the 6.5 to 6.7\,A and 30 to 120\,s range, respectively). The
temperature was kept constant at 20\,K for the duration of the experiments to maintain the most stable
experimental conditions before, during, and between subsequent potassium evaporations.

\section{Author Information}

The authors declare no competing financial interests. Correspondence and requests for materials should be
addressed to D. Fournier (dfournie@physics.ubc.ca) and A. Damascelli (damascelli@physics.ubc.ca).

\section{Acknowledgments}

We gratefully acknowledge W.A. Atkinson, J.P. Carbotte, D. Munzar, M.R. Norman, G.A. Sawatzky, T. Senthil,
and D. van der Marel for discussions. This work was supported by the Killam Program (A.D.), the Alfred P.
Sloan Foundation (A.D.), the CRC Program (A.D.), NSERC, CFI, CIFAR Quantum Materials, and BCSI. The Advanced
Light Source is supported by the Director, Office of Science, Office of Basic Energy Sciences, of the U.S.
Department of Energy under Contract No. DE-AC02-05CH11231.

\section{Supplementary Information}

\subsection{Effective hole-doping determination}

\begin{figure}[b!]
\includegraphics[width=0.95\linewidth]{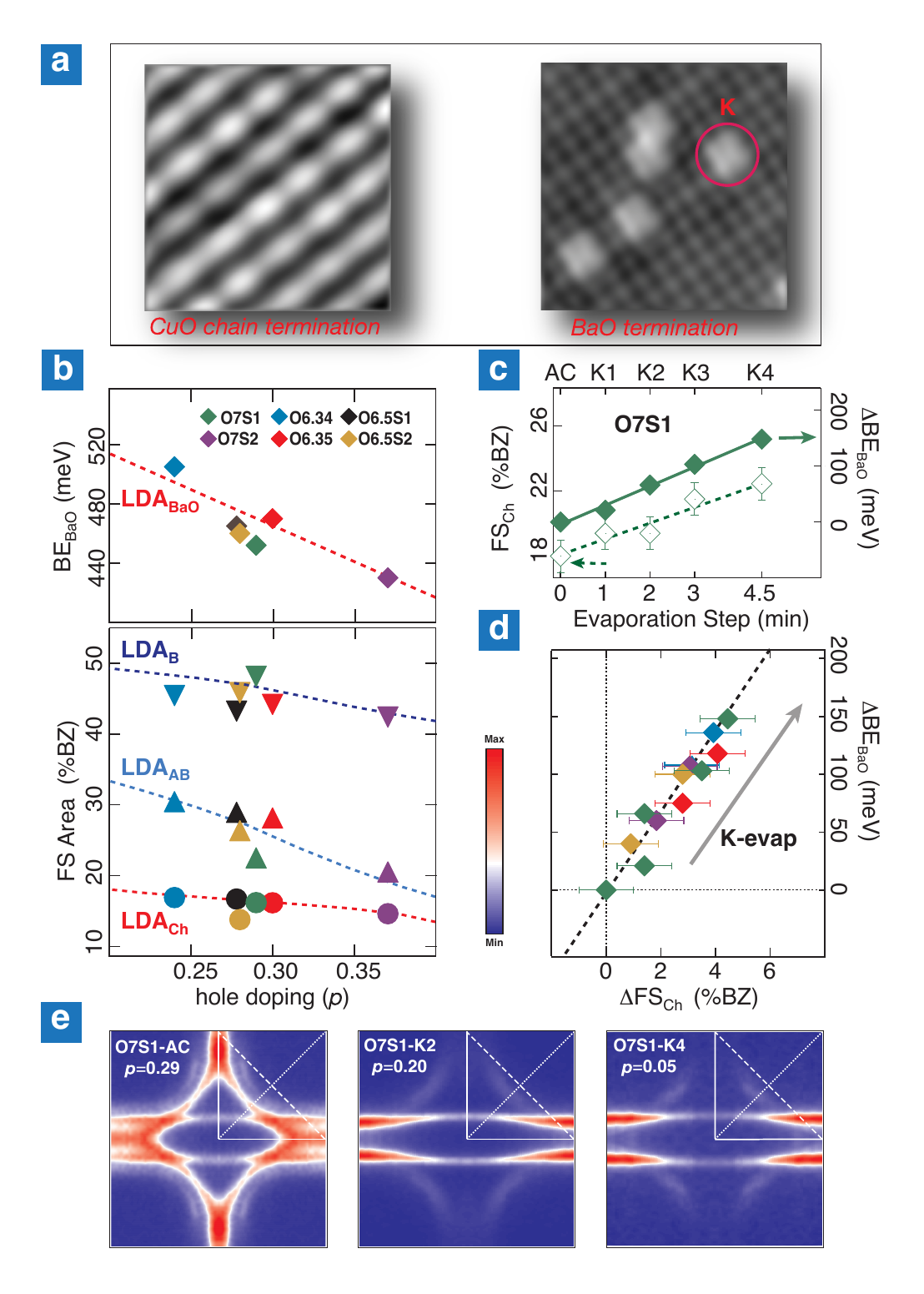}
\caption{{\bf K-evaporation: effective hole-doping determination.} (a) Topographic STM images (4x4\,nm$^2$) from oxygen-ordered ortho-II YBCO6.5
confirm the approximately 50-50\,\% CuO-BaO termination of the cleaved surface, the ortho-II alternation of full and empty chains, and the
preferential location on the BaO termination of evaporated K (red circle). (b) Binding energy of the BaO band ($BE_{BaO}$), and volume (counting
electrons) of $FS_{Ch}$, $FS_{B}$, and $FS_{AB}$, plotted versus the hole doping $p$ together with the results of LDA band-structure
calculations \cite{elfimov}. In this very overdoped regime, in which all Fermi surfaces are well-defined on both as-cleaved and moderately
K-deposited YBCO, the hole doping $p$ is determined from the average volume of $FS_{B}$ and $FS_{AB}$ (symbols refer to different bulk oxygen
contents). (c) Volume of $FS_{Ch}$ and variation of the BaO binding energy ($\Delta BE_{BaO}$) versus the K-evaporation time for YBCO7. (d) The
relative evolution of $\Delta BE_{BaO}$ and $\Delta FS_{Ch}$ upon K-deposition is linear over the whole range for all samples. (e) Fermi surface
mapping from as-cleaved and K-deposited YBCO7.}\label{Kdoping}
\end{figure}

As discussed in the paper, the data presented in Fig.\,1 and\,2 (paper), consistent with the observed doping and the $T\!=\!20$\,K temperature
of the experiments, reproduce the major features of the cuprate electronic structure in the overdoped normal metal ($p\!=\!0.29$, $T\!>\!T_c$),
the nearly-optimally doped superconductor ($p\!=\!0.20$, $T\!<\!T_c$), and the underdoped pseudogap state ($p\!=\!0.06$, $T_c\!<\!T\!<\!T^*$).
Most importantly, they demonstrate that in situ potassium doping of the cleaved surface of YBCO is a very effective mean of exploring the whole
cuprate phase diagram via surface-sensitive single-particle spectroscopies. To this end, a crucial aspect of the data analysis is the conversion
from K-evaporation to the effective hole doping $p$ per planar copper. If each K atom deposited on as-cleaved YBCO donates one electron to the
surface electronic structure, the resulting $K^{1+}$ ions should be more likely located on the insulating and charge-neutral BaO termination,
rather than on the (1+)-charged CuO chain termination. This is confirmed by our scanning tunneling microscopy (STM) study of various YBCO
samples after K-evaporation, which consistently finds K atoms on BaO terminations only; this is shown in Fig.\,\ref{Kdoping}a for the case of
ortho-II YBCO6.5. STM was performed at 10\,K and 2.0x10$^{-11}$\,torr with a beetle type STM, and the same SAES getter source and procedures
used in the ARPES experiments; tungsten tips were cleaned in situ by Ar sputtering and sharpened through gentle contact with an Au surface.
Topographic images on BaO (CuO chain) terminations were acquired with bias voltage and tunnelling current of $-200$\,mV and 200pA ($-600$\,mV
and 500\,pA). We studied a total of 10 samples of different bulk doping from the same batches as those used for ARPES.

With respect to the doping evolution of the top-most BaO, CuO-chain, and CuO$_2$-plane layers, the ansatz to be proven is that the electrons
donated by potassium will be sufficiently delocalized to lead to a uniform carrier doping. In Fig.\,\ref{Kdoping}b we show, for various
as-cleaved and lightly K-deposited samples, the binding energy of the BaO band ($BE_{BaO}$), as well as the volume (counting electrons) of
CuO-chain and (bonding and antibonding) CuO$_2$-plane Fermi surfaces ($FS_{Ch}$, $FS_{B}$, and $FS_{AB}$).
Note that, in this very overdoped regime in which all Fermi surfaces are well-defined, the effective hole doping $p$ can be determined directly
from the average volume of $FS_{B}$ and $FS_{AB}$, with no ambiguity. Interestingly, the same quantities estimated from LDA calculations for
fully oxygenated YBCO7, upon a rigid shift of the chemical potential, show good overall quantitative agreement suggesting that, beyond
$\simeq\!0.25$, the electronic structure is well described by independent-particle density-functional theory (for the BaO band, which LDA puts
closer to the Fermi energy, the values in Fig.\,\ref{Kdoping}b were obtained by shifting it down {\it self-consistently} to match the BE
observed for $p\!=\!0.37$ as the extreme starting point). As we further hole-underdoped the system below $p\!\simeq\!0.25$ by K-evaporation, we
cannot rely on the volume of $FS_{B}$ and $FS_{AB}$ in determining $p$ because of the opening of the superconducting gap and, eventually, the
collapse of the CuO$_2$ Fermi surfaces into disconnected Fermi arcs. We can however follow very clearly the increase of the (electron) $FS_{Ch}$
volume (Fig.\,\ref{Kdoping}e for YBCO7, sample 1) and of the BaO-band binding energy [Fig.\,1d,f,h (paper), for various bulk dopings]. The
results for all evaporation steps on YBCO7S1 are summarized in Fig.\,\ref{Kdoping}c, and demonstrate that both $FS_{Ch}$ and $BE_{BaO}$ scale
linearly with the amount of deposited potassium, all the way to the highest surface doping. Similar scalings are observed for all bulk oxygen
contents, leading to one universal linear scaling relation between $\Delta FS_{Ch}$ and $\Delta BE_{BaO}$, for all YBCO samples
(Fig.\,\ref{Kdoping}d). Note that the increase in both $FS_{Ch}$ and $BE_{BaO}$ upon K-evaporation (i.e., upward slopes in
Fig.\,\ref{Kdoping}c,d) confirms once again the electron-nature of the doping induced by K (i.e., effective hole-underdoping). Furthermore, the
linearity over the whole doping range is consistent with the uncorrelated behaviour predicted for both BaO and CuO chain-derived electronic
states in LDA+U calculations \cite{elfimov}. Quantitatively, the K-induced rate $\Delta BE_{BaO}/ \Delta FS_{Ch}\!=\!0.0323 \pm 0.0012 \, [eV \,
(\%BZ)^{-1}]$ obtained from Fig.\,\ref{Kdoping}d matches almost perfectly the ratio $\sim\!0.0331$ obtained from the evolution of $\Delta
BE_{BaO}$ and $\Delta FS_{Ch}$ versus hole doping $p$ in both as-cleaved results and LDA calculations in Fig.\,\ref{Kdoping}b [here we observed
$\Delta BE_{BaO}/ p \!=\!0.5059 \, (eV \, p^{-1})$ and $\Delta FS_{Ch}/ p \!=\!15.269 \, (\%BZ \, p^{-1})$]. This provides a uniquely accurate
measure of the charge carriers injected into the system upon K-evaporation, which can be expressed in terms of the usual hole doping parameter
$p$, i.e. the number of holes per planar copper away from half filling ($p\!=\!0$).

\subsection{Electrostatic effects from oxygen and potassium doping}

In the following we address the question of the sensitivity of bilayer splitting to doping and crystal structure for various oxygen
concentrations in bulk YBCO, as well as to the electrostatic effects associated with different O$^{2-}$ fillings in the CuO chain layers
sandwiching a CuO$_2$ bilayer, or the presence of the surface either as-cleaved or potassium evaporated. All these configurations have been
investigated by density functional theory (DFT) calculations using the linearized augmented-plane-wave method implemented in the WIEN2K code
\cite{wien2k}; exchange and correlation effects were treated within the local-density approximation (LDA) \cite{ldaPW}.
\begin{table}[t!]
\begin{center}
\begin{tabular}{c |  c} \hline \hline
 Structure/doping  & Nodal bilayer splitting (eV) \\
   \hline
Bulk YBCO$_{6.0}$ & 0.19  \\ \\
Bulk YBCO$_{6.5}$ & 0.21  \\ \\
Bulk YBCO$_{7.0}$ & 0.24  \\ \\

Slab YBCO$_{7.0}$  & \\
BaO$_{surf}$ & 0.18  \\
CuO$_{surf}$ & 0.23  \\ \\

Slab YBCO$_{7.0}$ with K  & \\
 BaO$_{surf}$ & 0.21 \\
 CuO$_{surf}$ & 0.19 \\
\hline \hline

\end{tabular}
\caption{Nodal bilayer splitting values for various YBCO bulk and surface structures, as obtained from WIEN2K density functional theory
calculations for $k_z\!=\!0$.}\label{ldaresults}
\end{center}
\end{table}
As a further check, we have also done calculations using the DFT code SIESTA \cite{Soler:2002p2745}, with the norm-conserving Troullier-Martins
pseudopotentials \cite{Troullier:1991p1993} and the double-$\zeta$ singly-polarized basis set; the exchange and correlation effects were again
treated within LDA, after Ceperley and Alder \cite{ldaCA}. The results obtained with the two methods are in very good agreement, with a
$\lesssim\!0.02$\,eV variation in bilayer splitting; the WIEN2K values are summarized in Tab.\,\ref{ldaresults}.

The YBCO$_{6+x}$ band structure calculations, for $x\!=\!1$ and $0$, were performed using the crystal structures reported by Jorgensen {\it et
al.} \cite{Jorgensen-prb-90} for $x\!=\!0.93$ and $0.09$, respectively. To study the effect of a non-symmetric potential on bilayer splitting in
a direction parallel to the c-axis, we calculated the electronic structure of YBCO6.5 using an artificial layered structure consisting of
alternating full and empty chain layers.
\begin{figure*}[t!]
\includegraphics[width=0.9\linewidth]{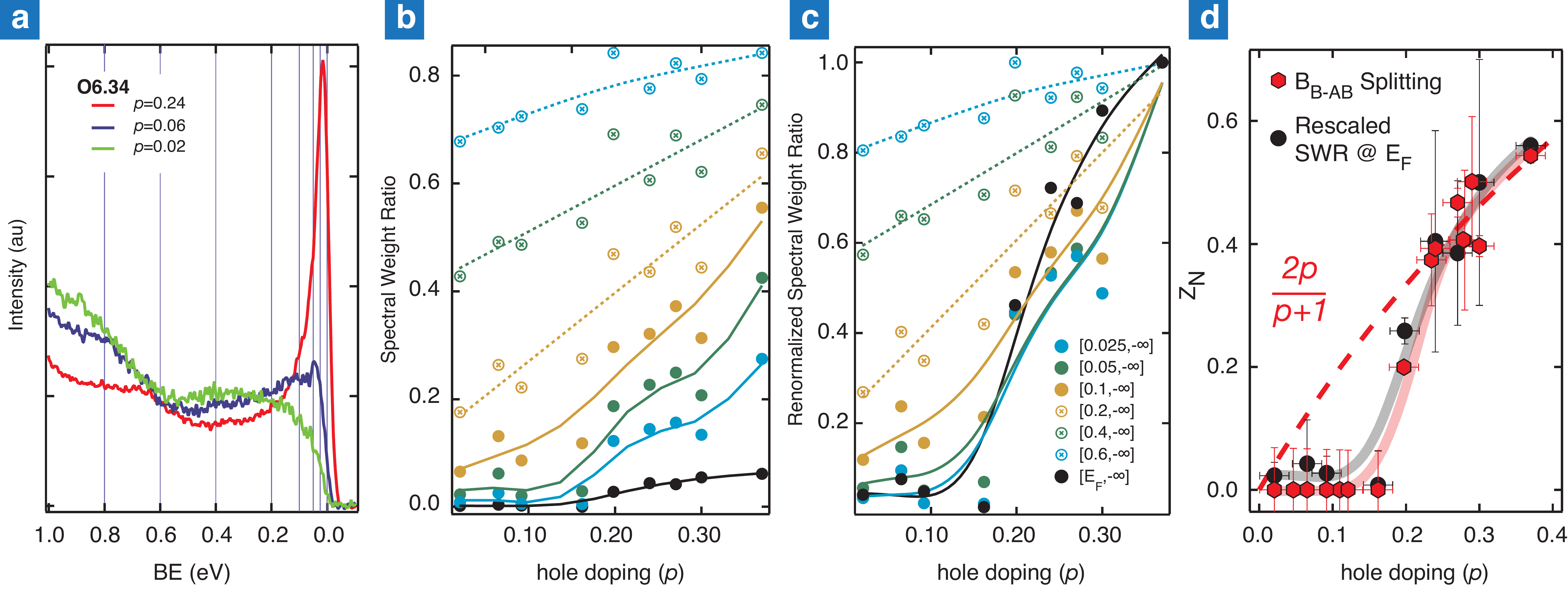}
\caption{{\bf Nodal quasiparticle integrity from spectral weight ratios.} (a) ARPES spectra at the nodal Fermi wavevector $k_{F,N}$ from
YBCO6.34. (b) Nodal spectral-weight ratio (SWR), defined as $\int^{-\infty}_{BE_{min}} I(k_{F,N},\omega)\, d \omega / \int^{-\infty}_{0.8\,eV}
I(k_{F,N},\omega)\, d \omega$, where $I(k,\omega)\!=\!I_0(k)f(\omega)A(k,\omega)$ is the intensity measured by ARPES; the data are presented
versus doping for progressively increasing integration windows (as indicated in c).  In (c), the same SWRs are renormalized to the $p\!=\!0.37$
value. (d) Nodal quasiparticle renormalization $Z_N$ as determined from the observed bilayer splitting and the SWR estimated at $E_F$ (the
latter was rescaled so that the two curves match in the $0.23\!-\!0.37$ doping range), together with the mean-field relation $2p/(p\!+\!1)$
\cite{andersonvanilla,yangrice}; thick gray and orange lines are guides to the eye. For the bilayer splitting data, the error bars are defined
from the B-AB MDC fits when bilayer splitting is detected, and from the experimental resolutions otherwise; for the SWR data, they are
calculated from the spread in the SWR values for integration windows smaller than $[0.1,-\infty]$ in binding energy (BE).}\label{Zintegral}
\end{figure*}
As one can see in Tab.\,\ref{ldaresults}, the bilayer splitting is only weakly dependent on doping. The influence on bilayer splitting of the
surface termination, either of BaO or CuO chain type, can be estimated with YBCO7.0 slab calculations. We adopted an ideal bulk crystal
structure and periodic slab geometry with 18\,\AA~separation between the slabs. The slabs are 3 unit cells thick and stoichiometric, with BaO
termination on one side and CuO chain termination on the another, which allows us to account for the effects associated with the polar
catastrophe \cite{mgo}. In the K-deposited YBCO7.0 slab calculations the K-position is relaxed, whereas the YBCO atomic positions are kept fixed
according to the bulk structure; the computed distance between K and BaO (CuO chain) surface is 2.57\,\AA\ (2.5\,\AA). As shown in
Tab.\,\ref{ldaresults}, also for `clean' and K-deposited BaO and CuO surface terminations we observe a negligible effect on the value of bilayer
splitting. We can thus conclude that, within DFT, the CuO$_2$ bilayer band splitting is almost independent of the doping as well as the change
in potential introduced by different filling of the CuO chain and the presence of the surface, whether as-cleaved or K evaporated. For the
purpose of a more direct comparison with the existing literature, in the paper we will refer to the values of nodal and antinodal bilayer
splitting $t_{\perp}^{LDA}(k)$ as obtained for YBCO7.0 in bulk calculations.

\subsection{Nodal quasiparticle weight estimate}

Formally, the degree of quasiparticle integrity is revealed by $Z_k\!\equiv\!\int\! A_{coh}(k,\omega)\,d\omega$, i.e. the integrated spectral
weight of the coherent part of the single-particle spectral function probed by ARPES \cite{Damascelli:RMP,Campuzano}. Experimentally, while in
the optimally-to-overdoped regime $Z_k$ is believed to be finite -- yet quantitatively undetermined -- at all momenta both above and below
$T_c$, the situation is much more controversial in the optimally-to-underdoped regime \cite{Damascelli:RMP,Campuzano}. In the following, as in
the rest of the paper, we concentrate on the behaviour of the nodal quasiparticles because they are better defined over a larger portion of the
phase diagram. In addition, since they are not affected by the opening of pseudo and superconducting gaps, they are not characterized by as
dramatic a temperature dependence as the antinodal quasiparticles; they can therefore be studied at low temperatures, while still providing
insights on the normal liquid underlying the emergence of HTSC. To illustrate the limitations even in the nodal region in the quantitative
estimate of the quasiparticle integrity from the spectral weight of $A_{coh}(k,\omega)$ as measured by ARPES, we present a detailed analysis in
Fig.\,\ref{Zintegral}. As shown in Fig.\,\ref{Zintegral}a, using YBCO6.34 as an example, a clear peak is detected at large doping values, which
loses intensity upon underdoping and eventually disappears. The first practical challenge in estimating $Z_k\!=\!\int\!
A_{coh}(k,\omega)\,d\omega$ is that ARPES does not measure directly $A(k,\omega)\!\equiv\!A_{coh}(k,\omega)\!+\!A_{incoh}(k,\omega)$ but rather
$I(k,\omega)\!=\!I_0(k)f(\omega)A(k,\omega)$. This is proportional to $A(k,\omega)$ via the Fermi-Dirac distribution function $f(\omega)$ and
the one-electron dipole matrix-element $I_0(k)$, which however remains experimentally undetermined (this is the so-called `matrix-element
effect' that makes the absolute value of the photoemission signal a strong and generally unknown function of polarization, photon energy, and
experimental geometry \cite{Damascelli:RMP,Campuzano}). To circumvent this problem, since the dipole matrix element is the same for both
$A_{coh}$ and $A_{incoh}$, one can try to evaluate experimentally $Z_k\!=\!\int I_{coh}(k,\omega)\, d \omega / \int I(k,\omega)\, d \omega$.
This is precisely the strategy followed for instance in Ref.\,\onlinecite{donglai} and \onlinecite{ding}, which however leads to two additional
challenges: (i) As already obvious from the data in Fig.\,\ref{Zintegral}a, $I_{coh}$ is strongly dependent on doping and progressively more
difficult to identify upon underdoping; a variety of phenomenological methods have been applied to estimate and remove the incoherent background
from $I(k,\omega)$, either based on the fit of the background by ad-hoc polynomial, linear, and integral functions \cite{donglai,ding}, or on
the estimate of the background from the spectral weight detected beyond $k_F$ \cite{kamibackground,plate} (i.e., from in principle unoccupied
regions of momentum space). (ii) $A_{incoh}$, and thus the integral of the complete $A(k,\omega)$, extend over a large energy scale given by
$U/2\!\simeq\!4$\,eV
 \cite{meinders,gabriel,elfimov}; they cannot be properly quantified due to the overlap with other electronic
states, such as oxygen bands starting already at 1-2\,eV BE in all cuprates and, in YBCO in particular, also BaO and CuO chain bands. As a
consequence, even if one could estimate $\int I_{coh}(k,\omega)\, d \omega $, an absolute quantitative value for $Z_k$ would still be lacking
due to $\int I(k,\omega)\, d \omega$.

To follow the evolution of the low-energy spectral weight versus doping, and attempt an heuristic distinction between its coherent and
incoherent components, in Fig.\,\ref{Zintegral}b we plot the nodal spectral weight ratio (SWR) defined as $\int^{-\infty}_{BE_{min}}
I(k_{F,N},\omega)\, d \omega / \int^{-\infty}_{0.8\,eV} I(k_{F,N},\omega)\, d \omega$, for various integration windows extending from above
$E_F$ ($-\infty$) to a finite binding energy below $E_F$ ($BE_{min}$). Although the SWR is numerically dependent on the arbitrary choice of the
full integration window ($[0.8,-\infty]$ in this particular case to avoid the large contribution from the oxygen-related features), it is
obvious that the narrower integration windows will capture more directly the degree of quasiparticle coherence. As one can see in
Fig.\,\ref{Zintegral}b, and even more clearly in Fig.\,\ref{Zintegral}c where the SWR data have been renormalized to the $p\!=\!0.37$ value,
there is a large systematic dependence on the choice of the integration window. Although it is not possible to reach any quantitative conclusion
based on such kind of data, a clear qualitative change occurs when the window extends beyond the 50\,meV BE range: for $BE_{min}\!\leq\!50$\,meV
the SWR consistently vanishes below $p\!\simeq\!0.15$, while it becomes finite at all doping -- and progressively larger -- for wider
integration windows (eventually approaching unity once $BE_{min}$ gets closer to 0.8\,eV). Since, as shown in Fig.\,\ref{Zintegral}a, the
sharpest component of the nodal spectra is located well within the first 50\,meV across the whole phase diagram, it is natural to associate the
low-energy SWR with a relative measure of the quasiparticle integrity $Z_N$; remarkably, the doping evolution of this low-energy SWR is also
reminiscent of the quasiparticle renormalization inferred from the observed bilayer splitting through the relation
$Z_N\!=\!\Delta\epsilon^{B,AB}_N/2 t_{\perp}^{LDA}(N)$. When the two results are plotted together (after the necessary rescaling of the SWR, so
that the two curves match in the $0.23\!-\!0.37$ doping range), the similarity is striking. This validates the quantitative determination of
$Z_N$ from the evolution of bilayer band splitting and confirms our conclusion of a rapid loss of nodal quasiparticle integrity in the
$p\!\simeq\!0.10\!-\!0.15$ doping range, possibly consistent with the results of cluster dynamical mean-field theory which predicts a rapid
suppression of $Z_N$ in this same doping range \cite{gabriel}.

\bibliography{YBCO_Z_CM}

\end{document}